\documentclass[11pt,a4paper]{article}
\usepackage{amsmath,amscd}
\usepackage{amssymb}
\usepackage{epsfig,subfigure,a4wide}
\usepackage{psfrag}

\begin{document}

 \title{Percolation fractal exponents without fractal geometry}
 \author{Agn\`es Desolneux$^1$ and Bernard Sapoval$^{1,2}$ \\
 $^1$ Centre de Math\'ematiques et de leurs
 Applications, CNRS, Ecole Normale Sup\'erieure, \\ 94235 Cachan, France\\
 $^2$ Laboratoire de Physique de la Mati\`ere
 Condens\'ee, CNRS, Ecole Polytechnique, \\
 91128 Palaiseau, France}

 \date{\today }
 \maketitle

\begin{abstract}
Classically, percolation critical exponents are linked to the power laws that 
characterize percolation cluster fractal properties.  It is found here that 
the gradient percolation power laws are conserved even for extreme gradient 
values for which the frontier of the infinite cluster is no more fractal. In 
particular the exponent 7/4 which was recently demonstrated to be the exact 
value for the dimension of the so-called "hull" or external perimeter of the 
incipient percolation cluster keeps its value in describing the width and 
length of gradient percolation frontiers whatever the gradient value. Its 
origin is then not to be found in the thermodynamic limit.  The comparison 
between numerical results and the exact results that can be obtained 
analytically for extreme values of the gradient suggests that there exist a 
unique power law from size 1 to infinity which describes the gradient 
percolation frontier, this law becoming a scaling law in the large system 
limit. 
\end{abstract}

%\pacs{ 64.60.Ak, 05.40.-a, 64.60.Fr}

Spreading of objects in space with a gradient of probability is most 
common. From chemical composition gradients to the distribution of plants 
which depends of their solar exposure, probability gradients exist in many 
inhomogeneous systems. In fact inhomogeneity is a rule in nature whereas most 
of the systems that physicists are studying are homogeneous as they are 
thought to be more simple to understand. In particular phase transitions or 
critical phenomena are studied in that framework, the simplest being 
percolation transition\cite{Stauffer}. 

In this work, the opposite situation, a strongly inhomogeneous system is
studied. We report the discovery that the validity of some percolation
exponents can be extended to situations which are very far from the large
homogeneous system limit. This is found in the frame of gradient percolation,
a situation first introduced in the study of diffusion front\cite{Sapo1,gp}.
Surprisingly these exponents, up to now believed to pertain to large systems,
are also verified in a limit that could be called the small system limit
(S.S.L.) where some of the gradient percolation properties can be computed
analytically. This letter presents a report concerning the 2D square
lattice. The same results and discussion have been obtained for the triangular
lattice and will be published elsewhere, together with the details of the
analytical calculations for both lattices. 

\begin{figure}
\includegraphics[width=0.7\columnwidth]{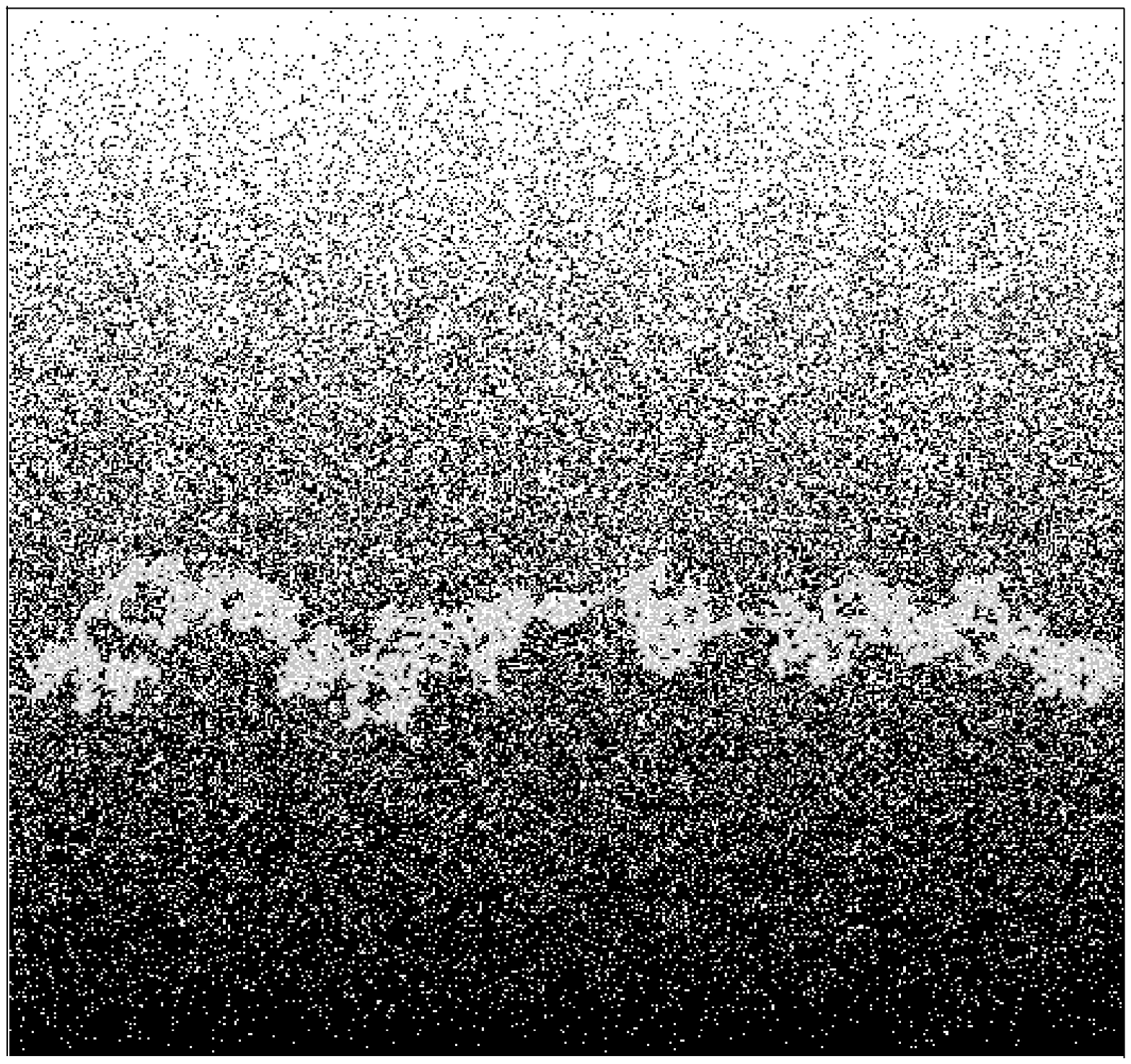}\\
\includegraphics[width=0.7\columnwidth]{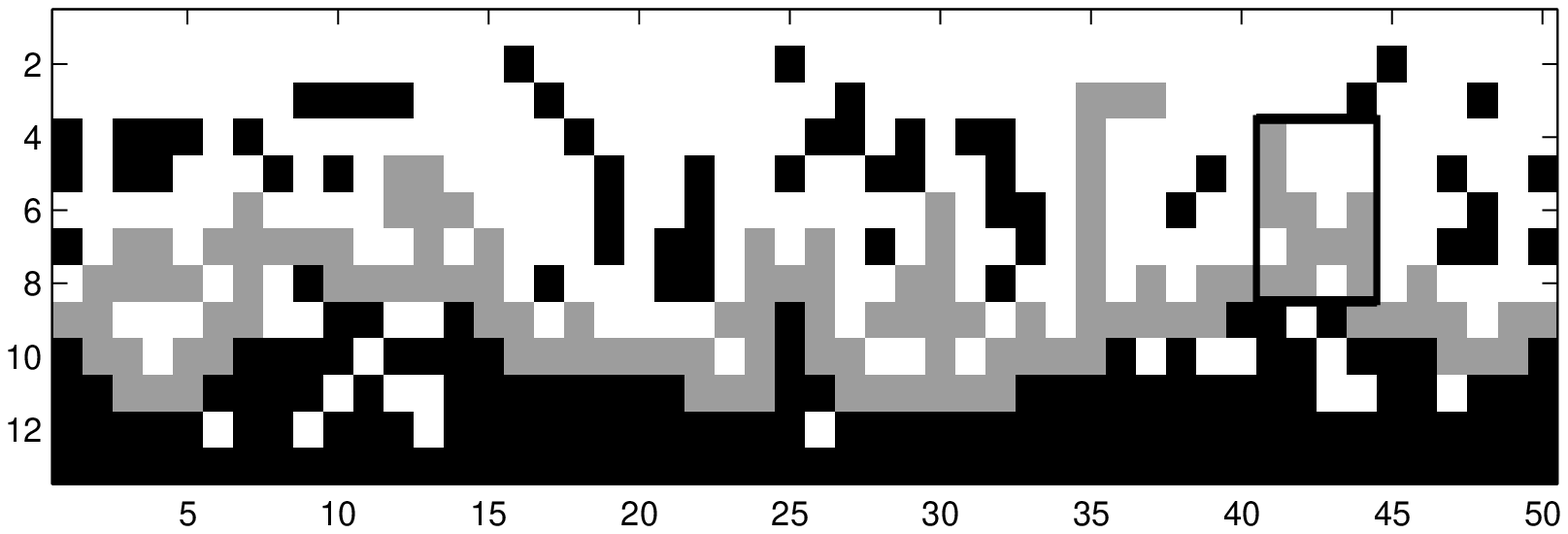}
\caption{ \label{fig:G.P.} Gradient percolation front. Particles are
distributed at random with  probability $p(x)=1-x/L_g$. The occupied 
sites are in black and the last line of connected occupied sites is the 
gradient percolation front shown in light gray. This situation corresponds
also to a diffusion situation where the front is called the diffusion
front. Top: $L=L_g=500$. Bottom: $L=50$ and $L_g=12$. The black window has an
horizontal width equal to $\sigma_f$. It contains approximately $L_g$ points.} 
\end{figure}

The gradient percolation (G.P.)  situation is depicted in
FIG.~\ref{fig:G.P.}. The figure gives an example of a random distribution of 
points on a lattice with a linear gradient of concentration in the vertical
direction.  It is a 2D square lattice of size $L_g\times L$, where each point
$(x,y)$ is occupied with probability $p(x)=1-x/L_g$ ($x$ being the vertical
direction in the figure). In gradient percolation there is always an infinite
cluster of occupied sites as there is a region where $p$ is larger than the
standard percolation (S.P.)  threshold $p_c$.  There is also an infinite
cluster of empty sites as there is a region where $p$ is smaller than $p_c$.
The object of interest is the G.P.  front, the external limit (or frontier) of
the infinite occupied cluster. It is constituted by the sites which belong to
the occupied cluster and are first-nearest neighbours with empty sites
belonging to the infinite empty cluster. It is shown in grey in
FIG.~\ref{fig:G.P.}.  This front is a random object with an average position
$x_f$, a statistical width $\sigma_f$ and a total length $N_f$. In so far that
the G.P. front and the S.P. external perimeter (often called hull) have the
same geometry their fractal dimension was first conjectured to be exactly
equal to 7/4 in \cite{Sapo1}.  This result was then demonstrated heuristically
by Saleur and Duplantier\cite{Saleur87} and very recently it was proved
mathematically by Smirnov\cite{Smirnov}. 

The early G.P. studies were focussed at finding its relation with standard
percolation. Let first recall the definitions. For $0\leq x\leq L_g$, $n_f(x)$
is the mean number of points of the front lying on the line $x$ per unit
horizontal length. It measures the front density at distance $x$. The length
$N_f$, the position $x_f$ and the width $\sigma_f$ of the front are then
defined in terms of the $n_f(x)$ by
$$N_f=L\sum_{x=0}^{L_g} n_f(x) , \hspace{0.2cm} x_f = \frac{\sum_{x=0}^{L_g} x
n_f(x)}{\sum_{x=0}^{L_g} n_f(x)} , $$
$$ {\mathrm{and}} \hspace{0.2cm} \sigma_f^2= \frac{\sum_{x=0}^{L_g}(x-x_f)^2
n_f(x)}{\sum_{x=0}^{L_g} n_f(x)} .$$

It was found that the mean front was situated at a distance where the density
of occupation was very close to $p_c$ or $p(x_f) \simeq p_c$. This was
verified numerically with such precision that  the gradient percolation method
is now often used to compute percolation
thresholds\cite{Rosso1,Rosso2,Ziff87,Quintinilla1,Quintinilla2}. It was also
found that: 
\begin{enumerate}
\item The width $\sigma_f$ depends on $L_g$ through a power law $\sigma_f
\propto (L_g)^{\nu/(1+\nu)}$ where $\nu=4/3$ is the correlation length
exponent \cite{Stauffer} in dimension $d=2$ so that $\sigma_f = (L_g)^{4/7}$.
The width $\sigma_f$ was also shown to be a percolation correlation length.
\item Secondly it was found that the front was fractal with a dimension $D_f$,
numerically determined, close to $1.75$. The front length followed a power law
$N_f \propto (L_g)^{\alpha_N}$ with $\alpha_N = (D_f-1)\nu/(1+\nu) $.
\item But also, it was numerically observed that the sum of these two
exponents was very close to 1. If true this meaned that $ \nu/(1+\nu) +
(D_f-1)\nu/(1+\nu) =1$ or $D_f = 1+1/\nu$ = 7/4.  This is how it was
conjectured in \cite{Sapo1} that $D_f = 7/4$.
\end{enumerate}
\noindent

In that sense the G.P. power laws were thought to be linked to the S.P.
exponent $\nu$ and to the fractality of the percolation cluster hull. Up to
now, these facts were considered to be strictly valid only in the large system
limit.

However, if true, and we know now that 7/4 is the exact value, there follows
an intriguing relation, namely $\sigma_f^{D_f}$ is exactly proportional to
$L_g^1$. This means that the number of surface particles within the
correlation length is exactly proportional to $L_g$. This is particularly
striking for diffusion fronts. Diffusion of particles from a source  results
in a concentration gradient and an associated G.P. situation. In that frame,
the above result means that, if $L_g$ particles have  diffused on a vertical
row, there is on average the same (or a constant fraction of) number of
particles on the correlated surface (surface content of a box with a lateral
size equal to the statistical width). This  fact seems a priori to have
nothing to do with scaling, percolation and the thermodynamic limit. From this
point of view it is possibly the consequence of a conservation law and if such
a conservation exists, it should apply also for extreme gradients
corresponding to $L_g$ of a few units. In particular it should apply to the
very extreme $L_g=1$, $2$ and $3$ for which exact values of $x_f$, $N_f$ and
$\sigma_f$ can be calculated analytically. 

For $L_g=2$ or $L_g=3$, given a site on a line $x$, one can describe all the
configurations such that the point belongs to the front, and  compute their
probability. For example, for $L_g=2$ all the  occupied  sites on the line
$x=1$ belong to the front, and a site on the line $x=0$ will belong to the
front  if at least one of its three neighbours on the line $x=1$ is
empty. Thus we get in this case, $n_f(1)=1/2$ and $n_f(0)=1-(1/2)^3=7/8$. With
the same kind of arguments, we can make the computations for $L_g=3$ but the
geometry of connected sets  is more complex as there are more configurations
to consider. This however can be done exactly.  One obtains: for $L_g=1$,
$\sigma_f =0$; \\  $\bullet$ for $L_g=2$:
$$ N_f/L = \frac{11}{8} , \hspace{0.2cm} x_f=\frac{4}{11} \hspace{0.2cm}
\mathrm{and} \hspace{0.2cm} \sigma_f = \frac{2\sqrt{7}}{11} .$$ $\bullet$ for
$L_g=3$:
$$ N_f/L = \frac{9401}{5832} , \hspace{0.2cm} x_f=\frac{6966}{9401}
\hspace{0.2cm} \mathrm{and} \hspace{0.2cm} \sigma_f =
\frac{9\sqrt{576049}}{9401}.$$

\begin{figure}
\includegraphics[width=0.7\columnwidth]{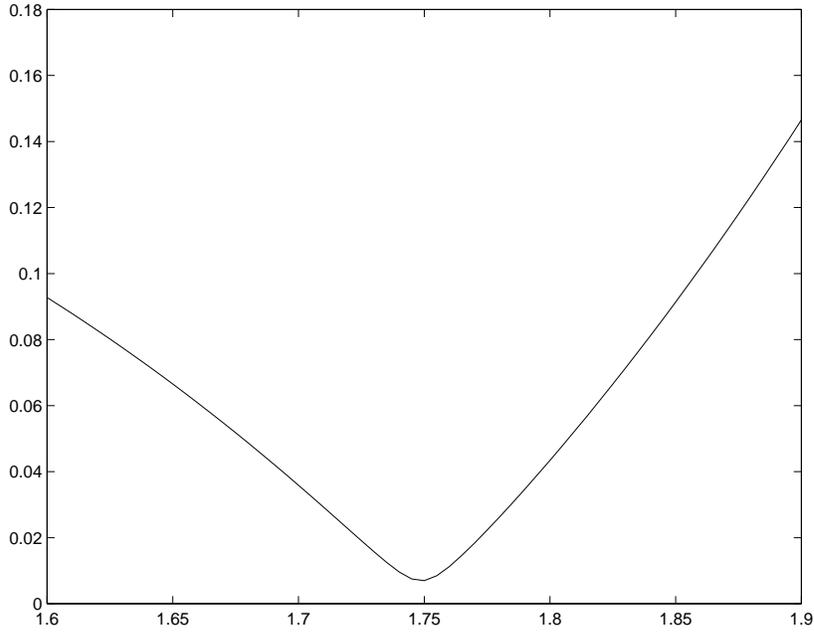}
\caption{\label{fig:1.75best} Determination of the best exponent value for the
square lattice: $\alpha=1.75$. The same result has been obtained for the
triangular lattice.}  
\end{figure}

The problem is then to compare the numerical G.P. laws to these exact
values. As will be shown, the numerical results verify the above power laws
with such precision that the question arises of the existence of a simple
mathematical power law extending from $L_g=1$ to infinity. To try to answer
this question we proceed in 2 steps. First we test these laws on the numerical
results obtained for $L_g$ between 4 and 50 for the square and triangular
lattices by searching the best numerical power laws followed by the
width. Considering arbitrary exponents $\alpha$ between $1.6$ and $1.9$, we
study $\sigma_f^{\alpha}$ as a function of $L_g$ between $4$ to $50$. For each
$\alpha$ value, there is a best line
$\sigma_f^{\alpha}=a_{\alpha}(L_g+b_{\alpha})$ fitting the numerical
$\sigma_f^\alpha$. The introduction of the term $b_{\alpha}$ is justified by
the fact that when a power law is verified for large systems it includes
always the possibility that a small (as compared to the system size) term
could contribute but in a negligible   manner. But here the  size itself is
small or very small. On the other hand, one should remark that for $L_g=1$ the
width is strictly 0 so that some negative value of $b_{\alpha}$ should be
present. In the next step, the  mean error $d(\alpha)$, defined by
$d(\alpha)^2=(1/47)\sum_{L_g=4}^{50}
(\sigma_f(L_g)^\alpha-a_{\alpha}(L_g+b_{\alpha}))^2$, is measured numerically
as a function of  $\alpha$. The results are shown in
FIG.~\ref{fig:1.75best}. There is a clear minimum for $\alpha=1.75$, showing
that this exponent gives the best power law fit. Once the best fit with the
empirical data is made one has the best values for the parameters $a$  and
$b$: $a=0.297$ and $b=-1.09$. Note that $b$ should be strictly equal to $-1$
in order to obtain a null width for the trivial case $L_g=1$.

\begin{figure} 
\includegraphics[width=0.7\columnwidth]{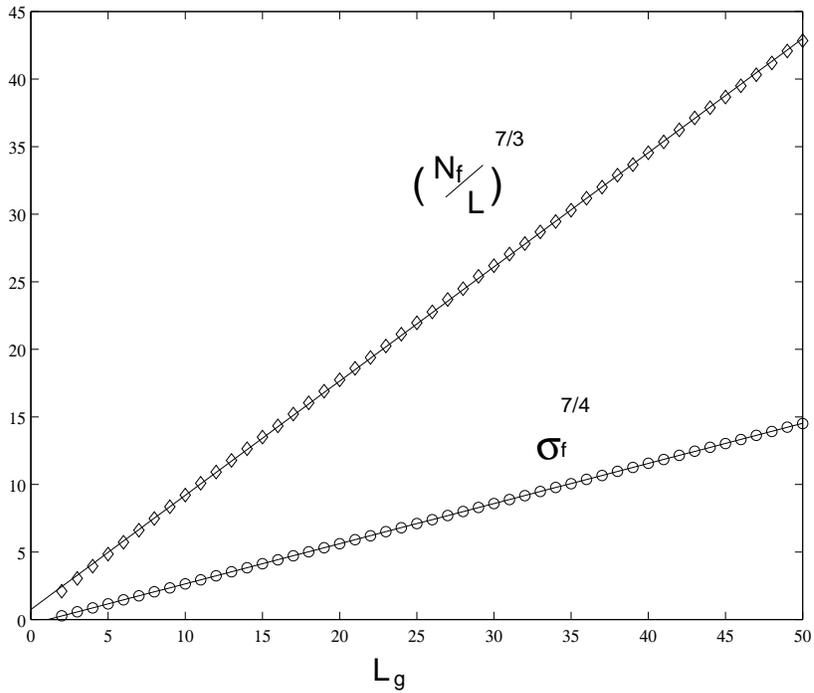}
\caption{ \label{fig:larg17etNf} Numerical results for the square lattice. The
circles resp. diamonds represent respectively $\sigma_f^{7/4}$ and
$(N_f/L)^{7/3}$. }
\end{figure}

An other verification of the extreme G.P. power laws can be obtained from
the study of the front length or of the quantity $(N_f/L)^{7/3}$ as a function
of $L_g$. In FIG.~\ref{fig:larg17etNf}, the diamonds represent the
 values of $(N_f/L)^{7/3}$ and the best linear fit has equation
$Y=c(L_g+d)$ with $c=0.845$ and $d=0.88$. This shows indeed that the exponents
$4/7$ and $3/7$ can be used down to the steepest gradients for which the
frontier is no more fractal.

One can then extrapolate the $\sigma_f$ values to  the case $L_g=1$, $2$ and
$3$. The results are given in Table I. One observes that the   numerical
extrapolations correspond to the exact values with an apparent good
precision. However no   firm conclusion can be  drawn without discussion of
the numerical uncertainties. The values shown  on FIG.~\ref{fig:larg17etNf}
are averaged over $50$ trials on a length $L=10^{5}$. As the fit occurs through
a power law it is difficult  to give a confidence interval for the
coefficients $a$ and $b$ (obtained from a least square linear regression   on
the values of $\sigma_f^{7/4}$). 

In order to obtain a better control on the numerical precision of $a$ and $b$
we made extensive  computations of the two cases $L_g=4$ and $L_g=5$ with 100
trials on a length $L=5.10^{5}$. Doing so  we obtain the mean values with
their standard deviation: $\sigma_f(4)^{7/4}=0.8658\pm 0.0009$ and for
$\sigma_f(5)^{7/4}=1.1610\pm 0.0013$.
Thus if we compute the equation of the line $\overline{a}(L_g+\overline{b})$
which interpolates the two points $(4,\sigma_f(4)^{7/4})$ and
$(5,\sigma_f(5)^{7/4})$, we obtain $\overline{a}=0.2952\pm 0.0022$ 
and  $\overline{b}=-1.066\pm 0.041$.
This last result shows that the value $-1$ is not incompatible with
$\overline{b}$ and its statistical error. Given the numerical values for
$L_g=4$ and $5$, we can also get extrapolated values for $\sigma_f$ for $L_g$
smaller, together with their confidence interval. The result (see Table I) is
that the predicted values are very close to the exact ones.  For
$(N_f/L)^{7/3}$, in the same way we obtain a linear interpolate of the values
for $L_g=4$ and $5$, with coefficients $\overline{c}=0,893\pm 0.014$ and
$\overline{d}=0.427\pm 0.074 $. 

At this point one could conclude that the numerical results are compatible
with the existence of a  single mathematical law for the width dependence,
this law working from $L_g=1$ to infinity. Note  however that the quality of
the random number generator should intervene. As long as a mathematical  proof
has not been given, it is thus impossible for us to conclude on the exact
values of the  coefficients $a$ and $b$.

The above discussion bears on the validity of the numerical results. But the
question of a unique  mathematical power law can also be studied from the
point of view of the exact results only. One can  note that, remarkably, the
values of the term $b$ in the fit of the width are close to $-1$. If the above
power law exists from $L_g=1$ to infinity  it suggests that the real value of
$b$ is exactly -1 as the width is null in the trivial case $L_g=1$. As we have
exact values, one can compute the  equation of the line $y=a(L_g+b)$ defined
by the two  points $(2,\sigma_f(2)^{7/4})$ and $(3,\sigma_f(3)^{7/4})$. One
obtains $a=0.294$ and $b=-1.055$. These values are close to the  values
obtained from the above numerical fit ($a=0.297$ and $b=-1.09$), and here
again $b$ is close but not equal to $-1$.

Why is there a small mismatch with the simplest law? The answer to this
question is two-fold. First,  it is possible that the exact power law is not
valid for $L_g=1$ or both $L_g=1$ and $2$. These cases could be "anormal" as
for these values there is no Grossman-Aharony  effect \cite{Grossman}. It is
then possible that these two cases do not enter the unique mathematical
law. Secondly the small discrepancy could be related to the fact that the
frontier definition  considers only the occupied sites. It gives to these
sites a privilege role whereas one should also  consider the frontier of the
empty cluster. In fact this is not new in percolation studies
\cite{Rosso1,Ziff87} where it was shown that the barycenter between the
frontier of the occupied cluster and the frontier of the empty cluster was a
more natural object. It notably permitted better computations of the
percolation threshold.  We have studied the statistical width of the local
barycenter which can also be computed exactly for   $L_g=2$ or $3$. The
results show the same behavior as described above i.e. a $b$ value close but
not   equal to $-1$. The question remains then open to define the nature of
the geometrical object which really would show a corresponding $b$ value equal
to $-1$. The answer has certainly interesting consequences for the S.P.
problem itself.

The above results have a simple practical consequence. Given an irregular
non-fractal interface, for instance the grey line shown in Fig. 1, one can
determine if it belongs to gradient percolation by measuring its statistical
width $\sigma$ and the average value of $N_f/L$. Through the results described
above one can find $L_g$ from the measured $\sigma_f$. Then one can check if
$N_f/L$ satisfies approximately the G.P. power law as a function of
$L_g$. This gives an intrinsic way to check if a given interface is of the
gradient percolation type {\it without previous knowledge} of the
gradient. This is important as there exist cases where irregular fronts like
corrosion fronts  belong to G. P. even though the gradient which built the
interface is no more present and only the  interface remains \cite{Balazs,GBS}.

In summary, it has been shown that the classical power laws  of gradient
percolation can  be extended to extreme gradients with the same fractal
exponents although the systems present no fractal geometry. Four comments can
be drawn on these results. First, extreme gradient situations can be found in
diffused contacts between materials. The fact that the contact geometry can be
described by the same set of exponent whatever the width of the diffused layer
is certainly of help to understand the properties of these contacts. There is
an intrinsic method to find if a given rough interface belongs to gradient
percolation without knowledge of the gradient. Secondly our results imply that
there exists a conservation law which stipulates that the length of the
correlated frontier is strictly proportional to the gradient length.  Thirdly,
the fact that the same exponent has been found for the  square and the
triangular lattice even for extreme gradients, in other words for small
systems, suggest that universality here is not related to the neglect of the
microscopic details of the interactions. Here there is no coarse scale and
still universality  is verified. Finally, the fact that the exponents 4/7 and
3/7 are valid down the smallest $L_g$ values (or the steepest gradients)
suggests that these exponents play the same type of role here that the
exponent 1/2 intervening in the fluctuations of the sum of independent
identical random variables. In that last case the exponent apply to {\it  any}
number of random variables starting from 1, 2 or 3 up to infinity.  The
exponents 4/7 and 3/7 may then play a stronger role that critical exponents
which only exist only in the thermodynamic limit.

The Centre de Math\'ematiques et de leurs Applications and the Laboratoire de
Physique de la Mati\`{e}re Condens\'{e}e are ``Unit\'{e} Mixte de Recherches
du Centre National de la Recherche Scientifique'' no. 8536 and ~7643.

\begin{table} 
\begin{center} 
\begin{tabular}{|c|c|c|c|} 
\hline  \, {\bf square lattice} \, & \, $L_g=1$ \, & \, $L_g=2$ \, & \,
$L_g=3$ \, \\  \hline  \, exact $\sigma_f$ \, & $0$ & $0.4810$ &  $0.7266$ \\
\hline  \,  $\sigma_f$ (4-50) \, & $-0.13$ & $0.47$ & $0.72$ \\  \hline  \,
$\sigma_f$ (4-5) \, & $0.106$  &  $0.478$  &  $0.726$ \\  \hline  \, $\delta
\sigma_f$ (4-5) \, & $0.023$  &  $0.005$  &  $0.002$ \\  \hline  \, exact
$N_f/L$ \, & $1$ & $1.3750$ &  $1.6120$ \\  \hline  \,  $N_f/L$ (4-50) \, &
$1.24$ & $1.48$ & $1.68$ \\  \hline  \,  $N_f/L$ (4-5) \, & $1.109$  &
$1.393$  &  $1.615$ \\  \hline  \, $\delta(N_f/L)$ (4-5) \, & $0.017$  &
$0.009$  &  $0.004$ \\  \hline 
\end{tabular} 
\medskip 
\caption{Comparison between exact and extrapolated results. The data $(4-50)$
resp. $(4-5)$ correspond to extrapolated values from the respective ranges
$(4-50)$ resp. $(4-5)$ (see text). $\delta$ is the confidence interval.} 
\end{center} 
\label{table1} 
\end{table}

\end{document}